\pdfoutput=1
\documentclass[11pt]{article}

\usepackage{acl}
\usepackage{times}
\usepackage{latexsym}
\usepackage[misc,geometry]{ifsym}
\usepackage[T1]{fontenc}
\usepackage[utf8]{inputenc}

\usepackage{microtype}

\usepackage[ruled]{algorithm2e}
\usepackage{subfigure}

\usepackage{microtype}
\usepackage{natbib}
\usepackage{caption}
\usepackage{graphicx}
\usepackage{amsmath}
\usepackage{color}
\usepackage{amsfonts}
\usepackage[switch]{lineno}
\usepackage{multirow}
\usepackage{tabu}
\usepackage{booktabs}
\usepackage[noend]{algpseudocode}
\usepackage{algorithmicx}
\usepackage{color}

 \newcommand{\nop}[1]{}
\title{Improving Continual Relation Extraction through\\Prototypical Contrastive Learning
\thanks{\quad This work is supported by Shanghai Science and Technology Innovation Action Plan (China) No.21511100401. 
}
}

\author{Chengwei Hu$^{1\dag}$, Deqing Yang$^{1\dag}$\textsuperscript{\Letter}, Haoliang Jin$^{1\ddag}$, Zhen Chen$^{1\ddag}$,
        \textbf{Yanghua Xiao$^{2\dag}$} \\
    $^1$School of Data Science, Fudan University, Shanghai, China  \\
    $^2$School of Computer Science, Fudan University, Shanghai, China \\
    $^{\dag}$\{cwhu20, yangdeqing, shawyh\}@fudan.edu.cn\\
    $^{\ddag}$\{hljin21, zhenchen21\}@m.fudan.edu.cn
  }

\begin{document}
\maketitle
\begin{abstract}
Continual relation extraction (CRE) aims to extract relations towards the continuous and iterative arrival of new data, of which the major challenge is the \emph{catastrophic forgetting} of old tasks. In order to alleviate this critical problem for enhanced CRE performance, we propose a novel \textbf{C}ontinual \textbf{R}elation \textbf{E}xtraction framework with \textbf{C}ontrastive \textbf{L}earning, namely \textbf{CRECL}, which is built with a classification network and a prototypical contrastive network to achieve the incremental-class learning of CRE. Specifically, in the contrastive network a given instance is contrasted with the prototype of each candidate relations stored in the memory module. Such contrastive learning scheme ensures the data distributions of all tasks more distinguishable, so as to alleviate the catastrophic forgetting further. Our experiment results not only demonstrate our CRECL's advantage over the state-of-the-art baselines on two public datasets, but also verify the effectiveness of CRECL's contrastive learning on improving CRE performance. %To reproduce our results conveniently, the related source codes and datasets are provided at https://github.com/PaperDiscovery/CRECL. 
\end{abstract}
\section{Introduction}\label{sec:intro}
%As one important task of information extraction and knowledge acquisition, relation extraction aims to identify the relation (class) between two entities mentioned in a sentence or one piece of text. It has played an essential role in many downstream applications including knowledge construction and completion \cite{liuqiao2016knowledge}.

In some scenarios of relation extraction (RE), massive new data including new relations emerges continuously, which can not be solved by traditional RE methods. To handle such situation, \emph{continual relation extraction} (CRE) \cite{wang2019sentence} was proposed. Due to the limited storage and computing resources, it is impractical to store all training data of previous tasks. %Thus the CRE models are required to extract the gradually emerging relations, that are different from traditional RE models only focusing on the limited data and relations.
%在实际场景中，面临着新的关系数据不断地，迭代地到来的场景，已有关系抽取的方法并不能很好的处理这个问题，因此continual relation extracti on被提出. 由于存储和计算资源有限，通常不太可能将所有之前已经见过的每类数据给存储下来，因此，CRE任务需要使用逐渐到来的新关系例子数据进行训练而不是像传统关系抽取任务那样有着固定的数据集和固定的关系数量。
As new tasks are learned where new relations emerge constantly, the model tends to forget the existing knowledge about old relations. Therefore, the problem of \emph{catastrophic forgetting} damages CRE performance severely \cite{hassabis2017neuroscience, thrun1995lifelong}. %The problem of catastrophic forgetting \cite{hassabis2017neuroscience, thrun1995lifelong} which refers to that old knowledge may be forgotten during the learning of new knowledge. In other words, 

In recent years, some efforts have focused on the alleviating catastrophic forgetting in CRE, which can be divided into consolidation-based methods \cite{kirkpatrick2017overcoming}, dynamic architecture methods \cite{chen2015net2net,fernando2017pathnet} and Memory-based methods \cite{chaudhry2018efficient,han2020continual,cui2021refining}. 
%the following three categories. 1) Consolidation-based methods \cite{kirkpatrick2017overcoming} adapt to each task by adjusting the parameter weights of each preceding task. 2) Dynamic architecture methods \cite{chen2015net2net,fernando2017pathnet} learn new tasks and alleviate old task forgetting by dynamically extending model structure. 3) Memory-based methods \cite{lopez2017gradient,aljundi2018memory,chaudhry2018efficient,cui2021refining} have attracted more attention recently, which employ an episodic memory module to memorize a small part of instances in old tasks to alleviate catastrophic forgetting.
%!!!!no incremental learning methods \cite{han2020continual,cui2021refining}
%wherein the memory-based methods have better CRE performance. Previous memory-based SOTA methods \cite{han2020continual,cui2021refining} assume that  all relations emerging in all tasks are known in advance, and achieve CRE task through classifying all instances into a fixed (relation) class set.} However, in many real-world scenarios we cannot know all new relations in advance, indicating that \emph{class-incremental learning \cite{masana2020class} is more expected than class-fixed learning in real-world CRE}. 
Despite these methods' effectiveness on CRE, most of them have not taken full advantage of the negative relation information in all tasks to alleviate catastrophic forgetting more thoroughly, result in suboptimal CRE performance.

%\red{Although some works \cite{wang2019sentence,wu2021curriculum} have adopted class-incremental learning mechanism, their neglect of negative sample information result in suboptimal performance.}
% and the inconsistence between model training and test

\begin{figure}[t]
\centering
%\vspace{-0.1cm}
\includegraphics[width=0.6\linewidth]{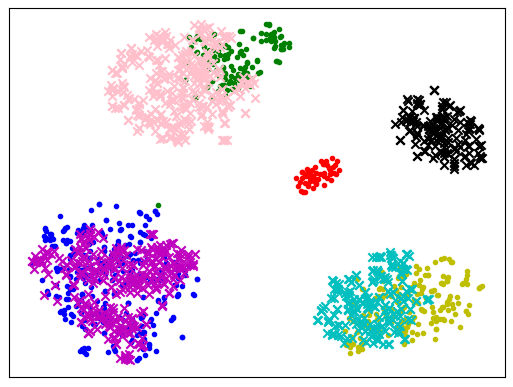}
\vspace{-0.2cm}
\caption{The data distribution map (better viewed in color) after training a classification model for an old task and then a new task. %The dots and crosses represent the instances of the old and new task, respectively, and different colors represent different relations. 
Many different relation data (different colors) of the old (dots) and new (crosses) task are mixed due to the catastrophic forgetting, making it hard to distinguish the new task's relations from the old task's relations.}
\label{fig:res1}
\vspace{-0.4cm}
\end{figure}
Through our empirical studies, we found that the catastrophic forgetting of a model results in the indistinguishability between the data (instances) distributions of all tasks, making it hard to distinguish the relations of all tasks. We illustrate it with the data distribution map after training a relation classification model for a new task, as shown in Figure \ref{fig:res1} where the dots and crosses represent the data of the old and new task respectively, and different colors represent different relations. It shows that the data points of different colors in either dot group (old task) or cross group (new task) are distinguishable. However, many dots and crosses are mixed, making it hard to discriminate the new task's relations from the old task's relations. Therefore, making the data distributions of all tasks more distinguishable is crucial to achieve better CRE. 
%The model suffer from such data distributions to distinguish the relations in the new task from that in the old task. 
%Such data distributions are due to the model's forgetting about the data distribution of old tasks when fitting new tasks.

To address above issue, in this paper we propose a novel \textbf{C}ontinual \textbf{R}elation \textbf{E}xtraction framework with \textbf{C}ontrastive \textbf{L}earning, namely \textbf{CRECL}, which is built with a classification network and a contrastive network. In order to fully leverage the information of negative relations to make the data distributions of all tasks more distinguishable, we design a \emph{prototypical contrastive learning} scheme. Specifically, in the contrastive network of CRECL, a given instance is contrasted with the prototype of each candidate relation stored in the memory module. Such sufficient comparisons ensure the alignment and uniformity between the data distributions of old and new tasks. Therefore, the catastrophic forgetting in CRECL is alleviated more thoroughly, resulting in enhanced CRE performance. In addition, different to the classification for a fixed (relation) class set as \cite{han2020continual,cui2021refining}, CRECL achieves an incremental-class learning of CRE which is more feasible to real-world CRE scenarios.
%CRECL's training process can be divided into three steps as follows. At first, for each relation in current task, we build a relation classification network to obtain the embeddings of all instances belonging to this relation. Then, from these instances we select some typical instances through K-means to represent this relation, which are stored in the memory module and subsequently used to generate the relation's prototype. On the second step, the instances in current task are compared with all (old and new) relations' prototypes which are obtained from the typical instances stored in the memory module. At last, for each typical instance stored in the memory module, we also compare it with the prototypes of all stored relations. Through such comparison between the (old and new) relation prototypes and instances, the data distributions of all tasks are adjusted to be more distinguishable, so as to alleviate the catastrophic forgetting more thoroughly. %As a result, the model improves the learning effects of current task without shrinking the learning effect of previous tasks.

Our contributions in this paper are summarized as follows: 

1. We propose a novel CRE framework CRECL that combines a classification network and a prototypical contrastive network to fully alleviate the problem of catastrophic forgetting. 

2. With the contrasting-based mechanism, our CRECL can effectively achieve the class-incremental learning which is more practical in real-world CRE scenarios.

3. Our extensive experiments justify our CRECL's advantage over the state-of-the-art (SOTA) models on two benchmark datasets, TACRED and FewRel. Furthermore, we provide our deep insights into the reasons of the compared models' distinct performance.

%The rest of this paper is organized as follows. We introduce previous research works related to ours in Section 2, and describe the details of our proposed CRECL in Section 3. Our experimental results are presented in Section 4. At last, we conclude our work in Section 5.

\section{Related Work}
% 可以从continual learning进展，contrastive learning进展，relation extraction进展
%Relation extraction as a basic task has widely used in many fields,
In this section, we briefly introduce continual learning and contrastive learning which are both related to our work. 

Continual learning \cite{clsurvey1,clsurvey2} focuses on the learning from a continuous stream of data. The models of continual learning are able to accumulate knowledge across different tasks without retraining from scratch. The major challenge in continual learning is to alleviate catastrophic forgetting which refers to that the performance on previous tasks should not significantly decline over time as new tasks come in. For overcoming catastrophic forgetting, most recent works can be divided into three categories. 1) Regularized-based methods impose constraints on the update of parameters. For example, LwF approach \cite{li2017lwf} enforces the network of previously learned tasks to be similar to the network 
of current task by knowledge distillation. However, LwF depends heavily on the data in new task and its relatedness to prior tasks. EWC \cite{ewc} adopts a quadratic penalty on the difference between the parameters for old and new tasks. It models the parameter relevance with respect to training data as a posterior distribution, which is estimated by Laplace approximation with the precision determined by the Fisher Information Matrix. WA \cite{zhao2020maintaining} maintains discrimination and fairness among the new and old task by adjust the parameters of the last layer. 2) Dynamic architecture methods change models' architectural properties upon new data by dynamically accommodating new neural resources, such as increased number of neurons. For example, PackNet \cite{packnet} iteratively assigns parameter subsets to consecutive tasks by constituting pruning masks, which fixes the task parameter subset for future tasks. DER \cite{yan2021dynamically} proposes a novel two-stage learning approach to get more effective dynamically expandable representation. 3) Memory-based methods explicitly retrain the models on a limited subset of stored samples during the training on new tasks. For example, iCaRL \cite{icarl} focuses on learning in a class-incremental way, which selects and stores the samples most close to the feature mean of each class. During training, distillation loss between targets obtained from previous and current model predictions is added into overall loss, to preserve previously learned knowledge. 
% Gradient Episodic Memory (GEM) \cite{gem} uses an episodic memory to store a subset of the observed samples from previous tasks. While minimizing the loss on current task, GEM treats the losses on the episodic memory as inequality constraints and avoids their increase. 
RP-CRE \cite{cui2021refining} introduces a novel pluggable attention-based memory module to automatically calculate old tasks' weights when learning new tasks.

\begin{figure*}
    \includegraphics[width =0.92 \linewidth]{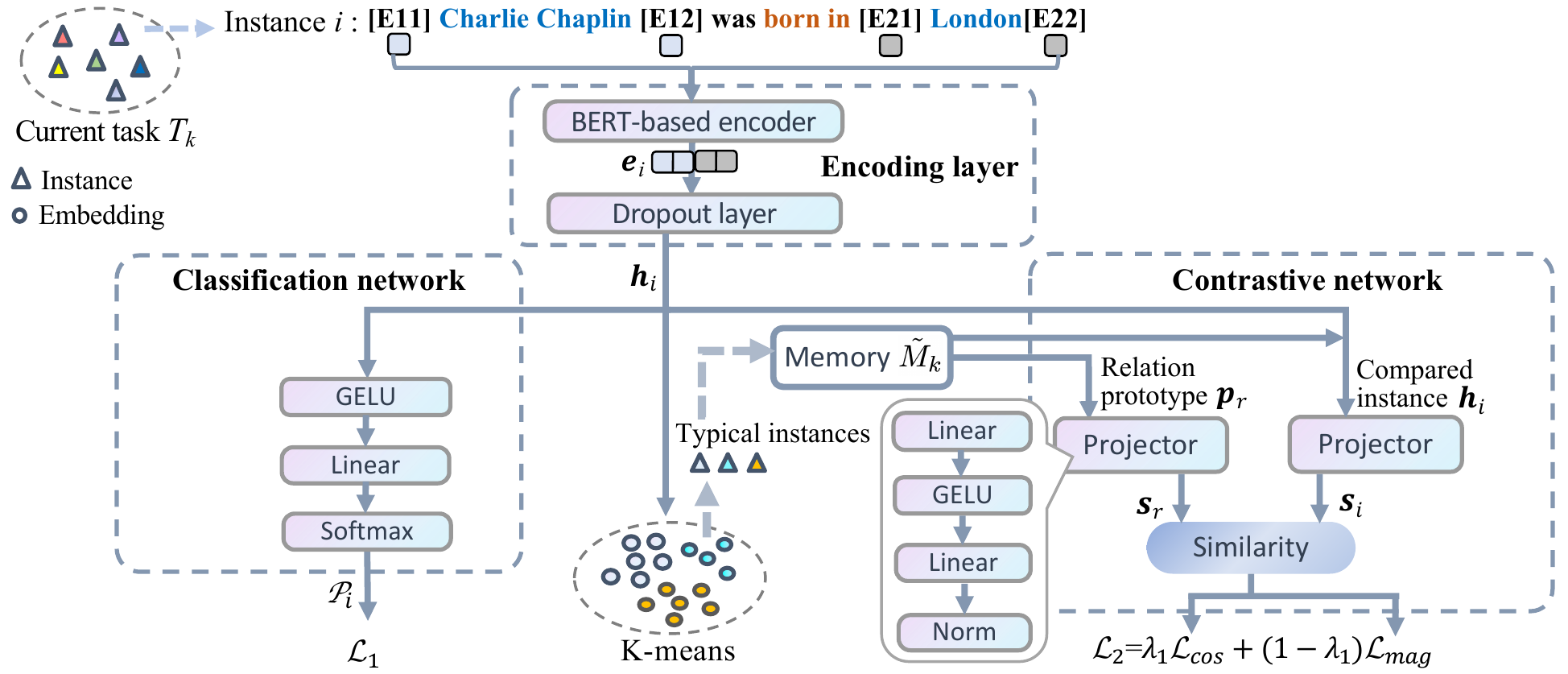}
   %\vspace{-0.1cm}
    \caption{The overall structure of our proposed CRECL. The framework is built with a shared encoding layer, a classification network and a contrastive network. %Solid lines indicate the need for gradient propagation, while dashed lines indicate the opposite.
    }
    \label{fig:framework}
    \vspace{-0.2cm}
\end{figure*}

% \paragraph{Contrastive Learning}
Since classification-based approaches require relation schema in the classification layer, classification-based models have an unignorable drawback on class-incremental learning. Many researchers leverage metric learning to solve this problem. \cite{wang2019sentence, wu2021curriculum} utilize sentence alignment model based on Margin Ranking Loss \cite{nayyeri2019adaptive}, while lack the intrinsic ability to perform hard positive/negative mining, resulting in poor performance. Recently, contrastive learning has been widely imported into self-supervised learning frameworks in many fields including computer vision, natural language processing and so on. Contrastive learning is a discriminative scheme that aims to group similar samples more closer and diverse samples far from each other. \cite{wang2021understanding} proves that contrastive learning can promote the alignment and stability of data distribution, and \cite{khosla2020supervised} verifies that using modern batch contrastive approaches, such as InfoNCE loss \cite{oord2018representation}, outperforms traditional contrastive losses, such as margin ranking loss, and also achieves good results in supervised contrastive learning tasks.

\section{Methodology}
    %\vspace{-0.1cm}
\subsection{Task Formalization}
    %\vspace{-0.1cm}
The CRE task aims to identify the relation between two entities expressed by one sentence in the task sequence. Formally, given a sequence of $K$ tasks $\{T_1,T_2,\ldots,T_K\}$, suppose $D_k$ and $R_k$ denote the instance set and relation class set of the $k$-th task $T_k$, respectively. $D_k$ contains $N_k$ instances $\{(x_1,t_1,y_1),\ldots,(x_{N_k},t_{N_k},y_{N_k})\}$ where instance $(x_i,t_i,y_i),1\leq i\leq N_k$ represents that the relation of entity pair $t_i$ in sentence $x_i$ is $y_i \in {R}_k$. 
% There are no identical relations in the $R_k$. 
One CRE model should perform well on all historical tasks up to $T_k$, denoted as $\tilde{{T}}_k =\cup_{i=1}^k {T}_i$, of which the relation class set is $\tilde{{R}}_k =\cup_{i=1}^k {R}_i$. We also adopt an episodic memory module ${M}_r = \{(x_1,t_1,r),\ldots,(x_L,t_L,r)\}$ to store typical instances of relation $r$, similar to \cite{han2020continual,cui2021refining}, where $L$ is the memory size (typical instance number). The overall episodic memory for the observed relations in all tasks is $\tilde{{M}}_k= \cup_{r \in \tilde{{R}}_k} {M}_r$. 

%\vspace{-0.1cm}
\subsection{Framework Overview}\label{sec:framework}
%\vspace{-0.1cm}
The overall structure of our CRECL is depicted in Figure \ref{fig:framework}, which has two major components, i.e., a classification network and a contrastive network. The procedure of learning the current task in CRECL is described by the algorithm in Alg. \ref{alg1}. 
%The left part is a relation classification network that uses special tokens to strengthen entities and uses regularized dropout to obtain an accurate encoder and prototype in this task. 

At first, suppose the current task is $T_k$, the representation of each instance in $T_k$ is obtained through the encoder and dropout layer shared by the two networks. In the classification network, each instance's relation is predicted based on its representation (line 1-3). Then, we apply K-means algorithm over the instance representations to select $L$ typical instances for each relation in $T_k$, which are used to generate the relation prototypes and stored into memory $\tilde{M}_k$ for the subsequent contrast (line 4-13).  
%The objective of this contrastive network is to adjust the data (instances) distributions of all tasks to be more distinguishable, which can alleviate the catastrophic forgetting in CRE as we claimed before. Therefore, it is necessary to serve for both current task and historical tasks, so 
There are two training processes in the contrastive network. The first is to compare current task instances with the stored relation prototypes of $\tilde{T}_k$ (line 14-17). The second is to compare each typical instance with all relation prototypes which are both stored in $\tilde{M}_k$ (line 18-24). These two training procedures ensure each compared instance keep distance from sufficient negative relations in $\tilde{R}_k$. Therefore, the data distributions of $\tilde{R}_k$ are distinguishable enough to alleviate CRECL's catastrophic forgetting of old tasks. Next, we detail the operations in CRECL.
%In the contrastive training process, the prototype remain unchanged while the instances still need to calculate gradients, which ensures the embeddings of the instances can be aware of stable targets.%It is worth noting that the encoder layer in the contrastive network and the encoder layer in the classification network are parameter sharing (the dotted box in Figure \ref{fig:framework}), while the parameter of the solid box in the contrastive learning is fixed as the value before the start of the contrastive training, which ensure that the generated prototypes are not disturbed by the training, and make the distribution more uniform through contrastive learning.
%右半边是一个基于原型和训练实例对比的对比学习网络，因为需要同时负责当前任务和历史任务，所以进行了两次训练，第一次是当前任务数据与原型进行的训练，旨在帮助对比学习模型快速学习到当前任务与原型的差异，第二次是memory中所存储数据与原型表示的对比训练，旨在帮助模型回顾之前任务从而减缓灾难性遗忘的发生。值得一提的是，对比学习网络中的encoder layer与分类网络中的encoder layer是参数共享的（图中虚线框），而对比学习中的实线框参数不参与更新，固定保持为对比训练开始前的值，这样做既可以保证生成的原型不会被对比学习训练所扰乱，也可以通过对比学习来调整使训练实例的嵌入的分布更为规整。

\begin{algorithm}[t]
\small
\caption{Training procedure for $T_k$}\label{alg1}
\LinesNumbered
%\KwIn{Training dataset $D_k$ of task k, current relation set $R_k$, history relation set $\tilde{R}_{k-1}$,history memory $\tilde{M}_{k-1}$,encoder $EN$, dropout Layer $DR$, classifier $CL$ projector $PJ$}
\KwIn{$D_k, R_k, \tilde{R}_{k-1}, \tilde{M}_{k-1}$}
%\KwOut{History memory $\tilde{M}_{k}$, encoder $EN$, dropout Layer $DR$} 
\KwOut{$\tilde{M}_{k}$} 
% $\tilde{R}_{k}$,sentence encoder \emph{\textbf{E}}, linear classifier \emph{\textbf{C}},  memory network \emph{\textbf{M}}
%\begin{algorithmic}[1]
%   $\mathcal{P}_k \gets \emptyset$\;
  \For{$i$ = 1 to epochs1}
{
     update Encoder $EN$, Dropout layer $DR$ and Classifier $CL$ by loss $\mathcal{L}_1$ on $D_k$\;
}
    $\tilde{R}_{k}=\tilde{R}_{k-1} \cup R_k$\;
    $\boldsymbol{p}=\emptyset$\;
  \For{ $r \in R_k$}
  {
    $d_r=\{d|d \in D_k,y_d = r$\}\; 
  	$\boldsymbol{h}_r$=DR(EN($d_r$))//Eq. 1\; 
%   	get $\boldsymbol{h}$ by Eq.\ref{eq:dropout}\;
  	apply K-means to all $\boldsymbol{h}_r$ and store $L$ typical instances into memory $M_{k}$\;
  	get prototype $\boldsymbol{p}_r$ from $M_{k}$ by Eq. 4\;
	$\boldsymbol{p}=\boldsymbol{p}\cup \boldsymbol{p}_r$\;
 }
  $\tilde{M}_{k}=\tilde{M}_{k-1} \cup M_k$\;
%  \For{each $r \in \tilde{R}_{k}$}
%  {
%     get $\boldsymbol{h}$ by Eq.\ref{eq:dropout}\;
%     get prototype $\boldsymbol{p}$ by average $h$ of relation $r$\;
%  }
    generate contrastive training data $C_k$ by random sampling from $D_k$\;
   \For{$i$ = 1 to epochs2}
{
     update $EN$, $DR$, $PJ$ by loss $\mathcal{L}_2$ on $C_k$ and $\boldsymbol{p}$\;
}
    generate contrastive training data $\tilde{C_k}$ by random sampling from $\tilde{M}_{k}$\;
\For{$i$ = 1 to epochs3}
{
    	\If{$y_i \in \tilde{C_k}$ is old class}
		{
		    use Eq. 1 to generate M embeddings\;
		}
     update $EN$, $DR$, $PJ$ by loss $\mathcal{L}_2$ on $\tilde{C_k}$ and $\boldsymbol{p}$\;
}
\end{algorithm}

% Next, we introduce the detailed operations in our CRE framework as follows.
%The following sections will give a detailed introduction to each part of our model in algorithmic flow order.
%接下来的几节会以算法流程顺序对我们模型的各个部分进行具体的介绍。
% 在本节中我们详细阐述了我们的框架整体的运行流程。正如图片\ref{fig:framework}、算法\ref{alg1}所示，我们的框架从结构上主要分为两个部分，从训练上来说会进行三次训练。

\subsection{Shared Encoding Layer}
%\vspace{-0.1cm}
The classification network and the contrastive network in CRECL are designed to promote each other, where the former classifies the current task based on its instance embeddings, and the latter effectively adjusts instance embeddings to keep uniformity and alignment. 
According to this principle, the two networks share the same layers in CRECL.

Specifically, for an instance $i$ of current task $T_k$, we use special tokens to represent the entities in $i$ as \cite{cui2021refining}. As shown in Figure \ref{fig:framework}, the head entity and tail entity in $i$ are represented by two special position tokens $[E11, E12]$ and $[E21, E22]$, respectively. The embedding of instance $i$ before the dropout layer, denoted as $\boldsymbol{e}_i\in\mathbb{R}^{2h}$, is the concatenation of token embeddings of $[E11, E12]$ and $[E21, E22]$ generated by BERT \cite{devlin2018bert} where $h$ is the dimension of two token embeddings. Then, $\boldsymbol{e}_i$ is fed into the dropout layer to obtain $i$'s hidden embedding as
\begin{equation}\label{eq:dropout}
    \boldsymbol{h}_i=\big(\boldsymbol{W}\operatorname{Dropout}(\boldsymbol{e}_i)+\boldsymbol{b}\big)\in \mathbb{R}^{d},
    %\vspace{-0.2cm}
\end{equation}
% \begin{equation}\label{eq:dropout}
%     \boldsymbol{h}=\operatorname{LN}\bigg(\operatorname{GELU}\big(\boldsymbol{W}\operatorname{Dropout}(\boldsymbol{e})+\boldsymbol{b}\big)\bigg)\in \mathbb{R}^{d},
% \end{equation}
where $\boldsymbol{W} \in \mathbb{R}^{{d}\times {2h}}$ (d is dimension of hidden layer) and $\boldsymbol{b}\in \mathbb{R}^d$ are both trainable parameters. In CRECL, $\boldsymbol{h}_i$ is regarded as $i$'s representation. 
% Here we emphasize dropout and linear transformation as dropout layer because in the next contrastive learning step, we use dropout layer to generate different embeddings for old instances.

% $\operatorname{GELU}(\cdot)$ is the activation function \cite{hendrycks2016gaussian}, and $\operatorname{LN}(\cdot)$ is the operation of layer normalization \cite{ba2016layer}. 
% encoder怎么做，dropout layer怎么做。
\subsection{Classifying Current Task}
%\vspace{-0.1cm}
With instance $i$'s representation $\boldsymbol{h}_i$, $i$'s probability distribution denoted as $\mathcal{P}_i\in \mathbb{R}^{|R_k|}$, is calculated in the classification network as
\begin{equation}\label{eq:output}
    \mathcal{P}_i=\operatorname{softmax}\big(\boldsymbol{W}_1\operatorname{LN}\big(\operatorname{GELU}(\boldsymbol{h}_i)\big)+\boldsymbol{b}_1\big),
        %\vspace{-0.1cm}
\end{equation}
where $\boldsymbol{W}_1 \in \mathbb{R}^{|R_k|\times {d}}, \boldsymbol{b}_1 \in \mathbb{R}^{|R_k|}$ are trainable parameters, and $|R_k|$ is the relation number of current task $T_k$ which is much less than the relation number of all tasks. $\operatorname{LN}(\cdot)$ is layer normalization operation. Then, classification loss for current task $T_k$ is calculated as
\begin{equation}\label{eq:enh}
    \mathcal{L}_1=
    -\frac{1}{N_k}\sum_{i=1}^{N_k}  \sum_{r=1}^{|R_k|} y_{i,r}\log \mathcal{P}_i^r,
      %\vspace{-0.2cm}
\end{equation}
where $y_{i,r}$=1 if $i$'s real relation label is $r$, otherwise $y_{i,r}$=0. $\mathcal{P}_i^r$ is the $r$-th entry in $\mathcal{P}_{i}$.

\subsection{Generating Relation Prototypes}\label{sec:proto}
%As we mentioned before, for each relation in current task, we select some typical instances to represent it which are stored in the memory module for the subsequent comparisons. To this end, 
After learning current task, for each relation $r$ in current task, we first apply K-means algorithm upon the representations ($\boldsymbol{h}_i$) of all instances belonging to $r$ to cluster them into $L$ clusters. Then, for each cluster, we select the instance most closest to the centroid of this cluster as one typical instance. Thus, $L$ typical instances of relation $r$ are selected and then stored into the memory module. With the stored typical instances of $r$, we average their representations as $r$'s \emph{prototype} $\boldsymbol{p}_r$, that is 
\begin{equation}\label{eq:prototype}
    \boldsymbol{p}_r=
    \frac{1}{L}\sum_{i=1}^{L}  \boldsymbol{h}_{i}^{r},
      %\vspace{-0.1cm}
\end{equation}
where $\boldsymbol{h}_{i}^{r}$ is a typical instance $i$'s representation of relation $r$. Such prototype best represents $r$ since the $L$ typical instances have the minimal distance sum to the $L$ cluster centroids. Another merit of such prototypes for representing relations is their insensitivity to the value of $L$. 

\subsection{Contrastive Network}
In this contrastive network, the instances are compared with the relation prototypes stored in the memory module to refine the data distributions of all tasks, so as to alleviate CRECL's catastrophic forgetting. Its basic principle is that, an instance's representation should be close to the prototype of its (positive) relation, and be far away from the prototypes of the rest (negative) relations. Please note that, the positive and negative relations are identified by the real labels of the training instances. Thus it is different from the self-supervised contrastive learning in other models \cite{chen2020simple}.

%Here we model contrastive learning as a comparison of a sentence with its prototype: the prototype itself is in the center of the class, and for those sentences that belong to the same class as the prototype, we think that it has similarities with the prototype, otherwise it should be treated as a negative sample. 

\paragraph{Contrastive Learning Objective}
As shown in the right part of Figure \ref{fig:framework}, the contrastive network is built with a twin-tower architecture. 
In the left tower, for a relation $r$, its prototypes $\boldsymbol{p}_r\in\mathbb{R}^{d}$ are obtained by Eq. \ref{eq:prototype}.
Then, $r$'s compared embedding is denoted as $\boldsymbol{s}_r\in \mathbb{R}^{\frac{d}{2}}$ and computed as
\begin{equation}\label{eq:project}
    \boldsymbol{s}_r=\boldsymbol{W}_3 \operatorname{GELU}(\boldsymbol{W}_2\boldsymbol{p}_r+\boldsymbol{b}_2)+\boldsymbol{b}_3,
% \vspace{-0.2cm}
\end{equation}
where $\boldsymbol{W}_2 \in \mathbb{R}^{{d}\times {d}}$, $\boldsymbol{b}_2 \in \mathbb{R}^{d}$, $\boldsymbol{W}_3 \in \mathbb{R}^{\frac{d}{2} \times {d}}$, $\boldsymbol{b}_3 \in \mathbb{R}^{\frac{d}{2}}$ are both trainable parameters.
\nop{
Then vector unitization is conducted
\begin{equation}\label{eq:regularity}
    \boldsymbol{s}^r=\frac{\boldsymbol{s}^r}{\Vert \boldsymbol{s}^r \Vert}.
 %\vspace{-0.2cm}
\end{equation}
}

In the right tower, for a compared instance $i$, its compared embedding is denoted as $\boldsymbol{s}_i\in\mathbb{R}^{\frac{d}{2}}$ and obtained by the same operation in Eq. \ref{eq:project} where only $\boldsymbol{p}_r$ is replaced by $\boldsymbol{h}_i$ from Eq. \ref{eq:dropout}.

For each instance $i$ in current task $T_k$, suppose the compared embedding of $i$'s relation $y_i$ is $\boldsymbol{s}_{y_i}$ which can also be obtained by Eq. \ref{eq:project}, since the typical instances of $y_i$ have been stored in $\tilde{M}_k$ before. We apply Euclidean norm to $\boldsymbol{s}_i$ and $\boldsymbol{s}_{y_i}$. Then, we use contrastive learning's InfoNCE loss \cite{oord2018representation} to calculate the cosine similarity loss of $T_k$ as
\begin{equation}\label{eq:cos}
    \mathcal{L}_{cos} =-\frac{1}{N_k}\sum_{i=1}^{N_k}\log \frac{\exp \left(\boldsymbol{s}_i  \boldsymbol{s}_{y_i} / \tau\right)}{\sum\limits_{r\in\tilde{R}_k} \exp \left(\boldsymbol{s}_i  \boldsymbol{s}_r/ \tau\right)},
%\vspace{-0.1cm}
\end{equation}
where $\tau$ is a temperature hyper-parameter.

To increase the similarity score gap of the correct label and the closest wrong label, inspired by \cite{koch2015siamese}, we propose a contrastive margin loss
\begin{equation}\label{eq:margin_loss}
    \mathcal{L}_{mag}=\frac{1}{N_k}\sum_{i=1}^{N_k} max\big( m- \boldsymbol{s}_{i}\boldsymbol{s}_{y_i}+\boldsymbol{s}_{i}\boldsymbol{s}_{k_i},0 \big),
%\vspace{-0.2cm}
\end{equation}
where relation $k_i=\mathop{\arg\max}\limits_{k\in\tilde{R}_k} \boldsymbol{s}_{i}\boldsymbol{s}_{k}$ s.t. $k \neq y_i$, that is $i$'s closest negative relation label. The margin loss penalizes that the similarity gap less than $m$. At last, the total loss is defined as
\begin{equation}\label{eq:contrast_loss}
    \mathcal{L}_{2}=\lambda_1\mathcal{L}_{cos}+(1-\lambda_1)\mathcal{L}_{mag},
%\vspace{-0.2cm}
\end{equation}
where $\lambda_1\in[0,1]$ is the controlling parameter.

\paragraph{Training Processes of Contrastive Learning}
There are two training processes in the contrastive network, %that follow the training process of the classification network introduced before. 
which both use the loss in Eq. \ref{eq:contrast_loss} to make the network parameters more fit to current task and all historical tasks, respectively.

% In the first training process, the $N$ in Eq. \ref{eq:contrast_loss} is set to $N_k$. It indicates that the network parameters are fine-tuned by contrastive learning with respect to (w.r.t.) the instances in current task $T_k$, and thus enhance the relation extraction performance of $T_k$. 
The first training process is conducted with current task $T_k$ and complements the classification network, it is an optional step with relatively small training epoch. However, it can not ensure the model fit to all tasks, because the model pays more attention to current task rather than the old tasks during this training process. In other words, as we have explained in the example of Figure \ref{fig:res1}, the model tends to ensure the instances of different relations in $T_k$ distinguishable, but forgets to meanwhile keep the instances of different relations in all historical tasks also distinguishable. As a result, the model's catastrophic forgetting still happens.

To alleviate CRECL’s catastrophic forgetting more thoroughly, we introduce the second training process in the contrastive network. 
% Specifically, we just set $N=L\times J$ in the loss of Eq. \ref{eq:contrast_loss}. 
% It indicates that 
In this process, all typical instances stored in the memory module are compared with all prototypes of stored relations, which cover in all tasks. 
We also conduct $M$ times forward propagation in the dropout layer, to generate $M$ embeddings for each old relation in $\tilde{R}_{k-1}$. Due to the randomness of dropout layer, we can get $M$ probability distributions for an old relation to reduce the imbalance of data distribution of old and new relation. 
Accordingly, this training process can effectively prevent the model from severe catastrophic forgetting. 
%The techniques used in the third training are the same as those used in the second training, except for samples. Using the representation of L typical samples stored in memory, training can be performed. Similarly, we consider the sentences in memory that belong to the same class as the prototype as positive samples, and the sentences in the memory that are different from the prototype as negative samples, because memory has stored data information of historical tasks in the previous steps, so training in this step can effectively prevent the model from catastrophic forgetting. 

% We have also tried to merge these two training processes. That is, both the stored typical instances and the instances of current task are used together as the compared instances. But due to the huge data gap, even if the focal loss \cite{lin2017focal} is used for balance, there is still a serious catastrophic forgetting problem. Thus we finally adopted the strategy of three separate training processes, i.e., the training in the classification network, followed by the two training process in the contrastive network. The previous work \cite{han2020continual,cui2021refining} also called this strategy as memory replay and reconsolidation.
%第三次训练所使用各项技术与第二次训练相同，除了采样的样本。使用在memory中存储下来的L个典型样本表示，就可以进行训练。类似的，我们将memory中与原型属于相同类的句子认为是正样本，将memory中与原型不同的类别句子认为是负样本，由于memory已经在前述步骤中存入了历史任务的数据信息，所以在这一步的训练可以有效地防止模型灾难性遗忘。我们也尝试了将第二次和第三次训练合并，即训练数据既包含历史数据，也包含当前任务的新数据，但由于数据量差距悬殊，哪怕使用了Focal loss进行平衡，在这种情况下依旧会发生严重的灾难性遗忘问题，因此我们使用了三次训练的方法，以往的工作\cite也把这种方法称为Memory Reconsolidation。

\subsection{Relation Prediction}
For a predicted instance $i$, we only measure its similarity to each stored relation, which is computed as the cosine distance between $i$'s representation and the relation's prototype. Then, we choose the most similar (closest) relation as $i$'s predicted class label, that is 
\begin{equation}\label{eq:pred}
    y^{*}_{i}=\mathop{\arg\max}\limits_{r \in \tilde{{R}}_k} \boldsymbol{s}_{i}\boldsymbol{s}_{r}.
%\vspace{0.2cm}
\end{equation}

% As we emphasized in Section \ref{sec:intro}, such contrasting-based mechanism ensures our framework's capability of class-incremental learning in the CRE.

\vspace{-0.2cm} 

\section{Experiments}
% In this section, we evaluate our CRE framework's performance through the comparisons with the SOTA CRE models.

\begin{table*}[t]
\centering
%\small
\caption{Accuracy (\%) comparisons on different test sets of historical cumulative tasks, showing that CRECL outperforms the compared models.}\label{table:overall}
\vspace{-0.2cm} 
\resizebox{0.78\textwidth}{!}{
\begin{tabular}{lllllllllll}
\hline

\hline
 & \multicolumn{10}{|c}{\textbf{FewRel}} \\ 
\cline{2-11}
\multicolumn{1}{l|}{\textbf{Model}}  & T1 & T2 & T3 & T4 & T5 & T6 & T7 & T8 & T9 & T10 \\ 
\hline
\multicolumn{1}{l|}{EA-EMR} & 89.0 & 69.0 & 59.1 & 54.2 & 47.8 & 46.1 & 43.1 & 40.7 & 38.6 & 35.2 \\
\multicolumn{1}{l|}{EMAR} & 88.5 & 73.2 & 66.6 & 63.8 & 55.8 & 54.3 & 52.9 & 50.9 & 48.8 & 46.3 \\
\multicolumn{1}{l|}{CML} & 91.2 & 74.8 & 68.2 & 58.2 & 53.7 & 50.4 & 47.8 & 44.4 & 43.1 & 39.7  \\
\multicolumn{1}{l|}{EMAR+BERT} &\underline{\textbf{98.8}}& 89.1 & 89.5 & 85.7 & 83.6 & 84.8 & 79.3 & 80.0 & 77.1 & 73.8 \\

\multicolumn{1}{l|}{RP-CRE+MA} & 98.0 & 91.4 & \underline{91.8} & 86.8 & 87.6 & \underline{86.9} & 83.7 & 81.9 & 80.1 & 79.5 \\
\multicolumn{1}{l|}{RP-CRE } & 97.9 & \underline{92.7} & 91.6 & \underline{89.2} & \underline{88.4} & 86.8 & \underline{85.1} & \underline{84.1} & \underline{82.2} & \underline{81.5}\\ 
\hline
\multicolumn{1}{l|}{CRECL+ATM(Ours) }&96.3 & 91.4 & 89.3 & 90.0 & 88.1 & 86.7 & 84.5 & 83.2 & 82.6 & 81.0 \\

\multicolumn{1}{l|}{CRECL(Ours)} & 97.8 & \textbf{94.9} & \textbf{92.7} & \textbf{90.9} & \textbf{89.4} &\textbf{87.5} & \textbf{85.7} & \textbf{84.6} & \textbf{83.6} & \textbf{82.7} \\ 
\hline
\multicolumn{1}{l|}{Improvement(\%)} & -1.01 & 2.37 & 0.98 & 1.91 & 1.13 & 0.69 & 0.71 & 0.59 & 1.70 & 1.47 \\ 
\hline

\hline
&\multicolumn{10}{|c}{\textbf{TACRED}} \\ 
\cline{2-11}
\multicolumn{1}{l|}{\textbf{Model}}  & T1 & T2 & T3 & T4 & T5 & T6 & T7 & T8 & T9 & T10 \\ \hline
\multicolumn{1}{l|}{EA-EMR} & 47.5 & 40.1 & 38.3 & 29.9 & 28.4 & 27.3 & 26.9 & 25.8 & 22.9 & 19.8 \\
\multicolumn{1}{l|}{EMAR} & 73.6 & 57.0 & 48.3 & 42.3 & 37.7 & 34.0 & 32.6 & 30.0 & 27.6 & 25.1 \\
\multicolumn{1}{l|}{CML} & 57.2 & 51.4 & 41.3 & 39.3 & 35.9 & 28.9 & 27.3 & 26.9 & 24.8 & 23.4  \\
\multicolumn{1}{l|}{EMAR+BERT} & 96.6 & 85.7 & 81.0 & 78.6 & 73.9 & 72.3 & 71.7 & 72.2 & 72.6 & 71.0 \\

\multicolumn{1}{l|}{RP-CRE+MA} & 97.1 & \underline{91.4} & \underline{87.4} & 82.1 & 78.3 & \underline{77.8} & 74.9 & 73.5 & \underline{73.6} & 72.3 \\
\multicolumn{1}{l|}{RP-CRE } & \underline{\textbf{97.6}} & 90.6 & 86.1 & \underline{82.4} & \underline{79.8} & 77.2 & \underline{75.1} & \underline{73.7} & 72.4 & \underline{72.4} \\ 
\hline
\multicolumn{1}{l|}{CRECL+ATM(Ours) } & 93.2& 80.2 & 77.3 & 76.0 & 71.8 & 71.5 & 69.2 & 72.3 & 70.0 & 71.2  \\
\multicolumn{1}{l|}{CRECL(Ours) } & 96.6 &\textbf{93.1} & \textbf{89.7} & \textbf{87.8} & \textbf{85.6} &\textbf{ 84.3} & \textbf{83.6} &\textbf{81.4} & \textbf{79.3} &\textbf{78.5} \\ 
\hline
\multicolumn{1}{l|}{Improvement(\%)} & -1.02 & 1.86 & 2.63 & 6.55 & 7.27 & 8.35 & 11.32 & 10.45 & 7.74 & 8.43 \\
\hline

\hline
\end{tabular}
}
\vspace{-0.2cm} 
\end{table*}

\subsection{Datasets}
Our experiments were conducted upon the following two benchmark CRE datasets.

\noindent\textbf{FewRel} \cite{han2018fewrel} is a popular relation extraction dataset originally constructed for few-shot relation extraction. The dataset is annotated by crowd workers and contains 100 relations and 70,000 samples in total. In our experiments, to keep consistent with the previous baselines, we used its version of 80 relations. 

\noindent\textbf{TACRED} \cite{zhang2017tacred} is a large-scale relation extraction dataset containing 42 relations (including \emph{no\_relation}) and 106,264 samples from news and web documents. Based on the open relation assumption of CRE, we removed \emph{no\_relation} in our experiments. To limit the sample imbalance of TACRED, we limited the number of training samples of each relation to 320, and the number of test samples of each relation to 40, which is also consistent with previous baselines. Compared with FewRel, the tasks in TACRED are more difficult due to its relation imbalance and semantic difficulty.

\subsection{Compared Models}
\vspace{-0.1cm}
We compare our framework with the following baselines in our experiments.

\noindent\textbf{EA-EMR} \cite{wang2019sentence} proposes a sentence alignment model with replay memory module to alleviate catastrophic forgetting. 

\noindent\textbf{EMAR} \cite{han2020continual} proposes a novel memory replay, activation and reconsolidation method to alleviate catastrophic forgetting effectively.  

\noindent\textbf{EMAR+BERT} is an advanced version of EMAR where the original encoder (Bi-LSTM) is replaced with BERT.

\noindent\textbf{CML} \cite{wu2021curriculum} proposes a curriculum-meta learning method to tackle the order-sensitivity and catastrophic forgetting in CRE.  

\noindent\textbf{RP-CRE} \cite{cui2021refining} is a SOTA CRE model introducing a novel pluggable attention-based memory module to automatically calculate the weight of old tasks when learning new tasks.

\noindent\textbf{RP-CRE+MA} is an advanced version of RP-CRE where a memory activation step is added before attention operation. %Similar to our framework, three training processes are required.

%In the following experiment figures and tables, our proposed framework is named as \textbf{CRECL} (Continuous Relation Extraction with Contrastive Learning). 
In our CRECL, we adopted the \textit{Bert-base-uncased} pre-trained by HuggingFace \cite{wolf-etal-2020-transformers} as the encoder, which is also used in {EMAR+BERT}, {RP-CRE} and {RP-CRE+MA}. Other baselines cannot be easily replaced by the BERT due to their architectures. In addition, we propose another version of CRECL, namely \textbf{CRECL+ATM}, which incorporates an attention memory module proposed by \cite{cui2021refining} in the contrastive network and used to verify its effectiveness of refining relation prototypes. 

% 为了较为方便的展示模型是否发生灾难性遗忘，我们使用T1～T10来表示历史累积任务，即在至T时刻出现过的所有任务的测试集，Accuracy (\%) on all observed relations at the stage of learning current task, indicating that our model (CRECL) significantly outperforms other models.

\begin{figure*}[htbp]
    \centering
    \subfigure[Accuracy of historical cumulative tasks on FewRel.]{
        \includegraphics[width=3.0in]{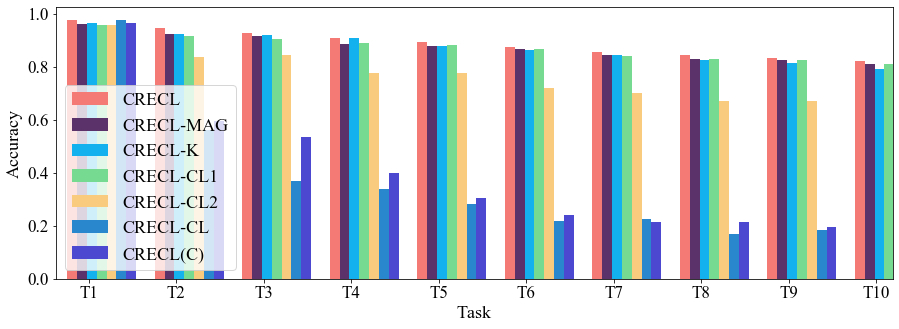}
    }
     \vspace{-0.4cm}
    \subfigure[Accuracy of current task on FewRel.]{
	\includegraphics[width=3.0in]{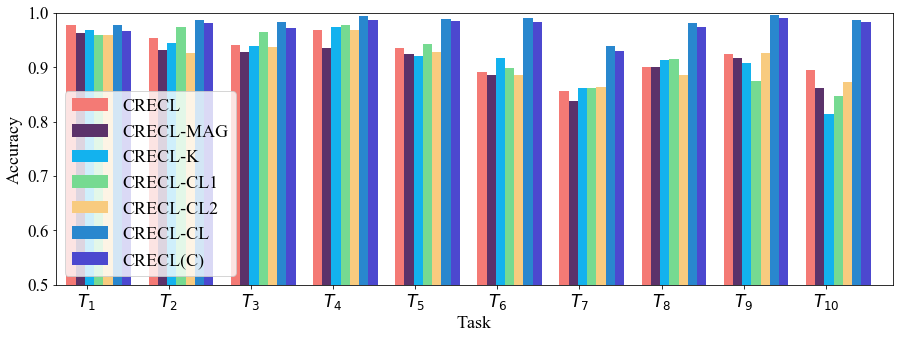}
    }
%     \quad    %用 \quad 来换行
    \subfigure[Accuracy of historical cumulative tasks on TACRED.]{
    	\includegraphics[width=3.1in]{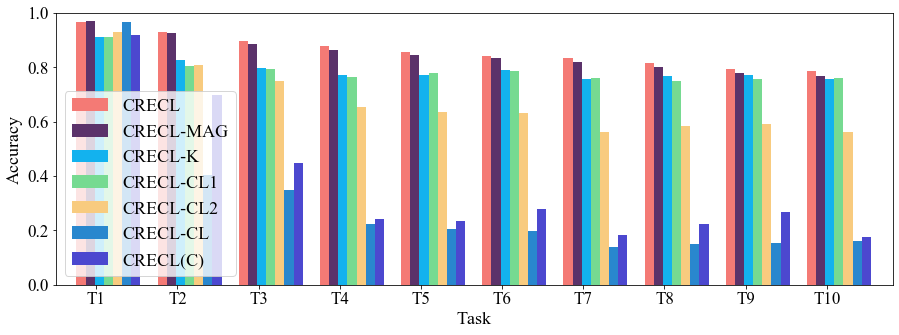}
    }
        \hspace{-0.4cm}
    \subfigure[Accuracy of current task on TACRED.]{
	\includegraphics[width=3.1in]{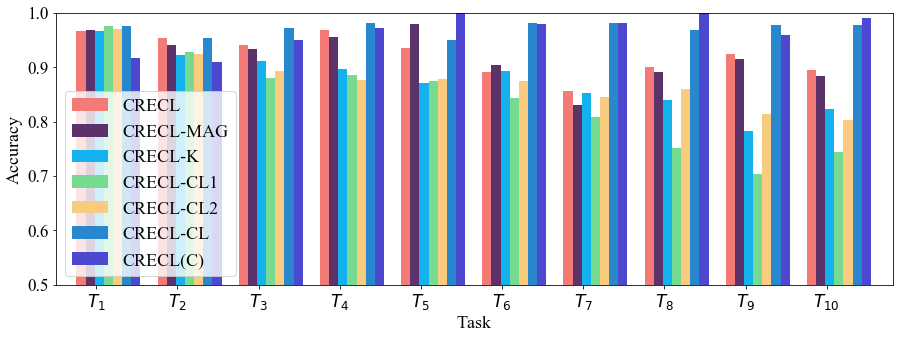}
    }
    \vspace{-0.4cm}
    \caption{Ablation study results on historical cumulative tasks and current task.}
    \label{fig:ablation}
    \vspace{-0.2cm}
\end{figure*}

\subsection{Experimental Settings}\label{sec:set0}
\vspace{-0.1cm}

Our evaluation metric is \textit{Accuracy} which is popularly used in previous baselines. 

For fair comparisons, we followed the experiment settings in RP-CRE. At first, to verify whether a CRE model suffers from catastrophic forgetting, we use $Tk$ to represent the test set of all historical cumulative tasks from the first task to the $k$-th task $T_k$ (Please note the difference between $Tk$ and $T_k$). In our ablation studies, we also report the performance on the test set of current task. To simulate different tasks, we randomly divided all instances into 10 groups (corresponding to 10 tasks). The task order of all compared models is exactly the same to reduce contingency. We also set the memory size in the baselines the same as ours. Relations are first divided into 10 clusters to simulate 10 tasks. All the reported results of the related baselines are the same as \cite{cui2021refining}.
% The rest settings are specified in Appendix \ref{sec:set}. 
For those special hyper-parameters in our experiments are as follows. The batch size is 32, the learning rate is set to 5e-5, $\tau$ is 0.08. We adopted 10 and 15 classification epochs for TACRED and FewRel, respectively. We also adopted 10 epochs for the first training process (for current task) and 5 epochs for the second training process (for all tasks) in the contrastive network. 

Because the total matrix operations and the data amount of second training in contrastive learning are very small, CRECLl's training time (1h31min) is very close to the SOTA model RP-CRE (1h28min).
%因为全矩阵化运算以及第三次训练的数据量非常少，我们的模型与只使用了二次训练的SOTA的训练时间非常接近，
To reproduce our experiment results conveniently, CRECL's source code together with the datasets are provided at \url{https://github.com/PaperDiscovery/CRECL}.
%对于那些独特的超参数来说，我们设置,所有我们的报告的实验至少重复了10次并报告了结果的平均值。

\subsection{Experimental Results and Analyses}
The following reported results of CRECL and its ablated variants are the average scores of running models for 5 times.
\vspace{-0.1cm} 
\subsubsection{Overall Performance Comparisons}
\vspace{-0.1cm} 
The overall performance of all compared baselines are reported in Table \ref{table:overall}, where the results of the baselines directly come from \cite{cui2021refining} and the baselines' hyper-parameter settings were the same as their original papers. The last row in the table is the improvement ratio of CRECL's performance relative to the best baseline's performance (underline). Based on these results, we have the following conclusions:

\begin{figure*}[!htbp]
    \centering
    \subfigure[CRECL]{
        \includegraphics[width=1.95in]{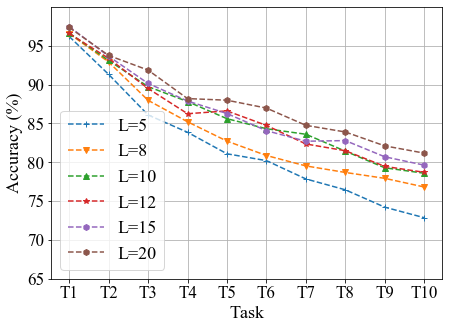}
    }
    \subfigure[RP-CRE]{
	\includegraphics[width=1.95in]{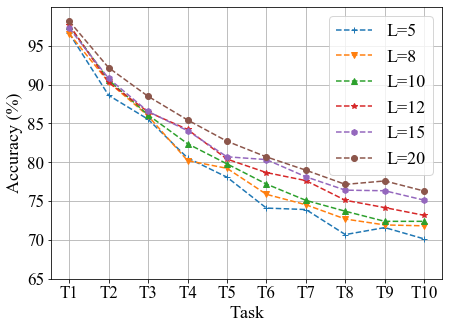}
    }
    \subfigure[CRECL vs. RP-CRE in T9/T10]{
	\includegraphics[width=1.95in]{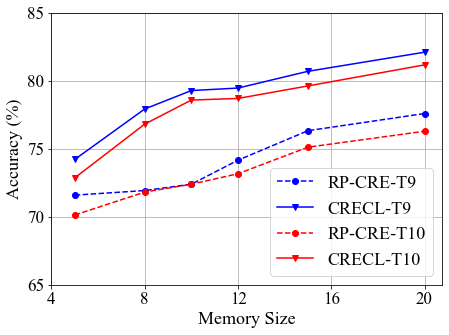}
    }
    \vspace{-0.2cm}
    \caption{Performance comparisons with different memory sizes on TACRED.}
    \label{fig:mem_diff}
        \vspace{-0.2cm}
\end{figure*}
(1) Our CRECL outperforms the SOTA model RP-CRE on both datasets. Compared with FewRel, CRECL has more apparent improvement over the baselines on TACRED. This may be due to that FewRel's tasks are not difficult enough as TACRED, proving that CRECL is good at handling more difficult tasks. %
% 1.我们的方法在两个数据集上都超过当前的SOTA RP-CRE。在fewrel上我们的方法有一些领先，而在TACRED数据集上我们的方法展现出了更为明显的改善，这可能是由于fewrel任务还不够难，而TACRED由于其类别不均衡性，并且句意更加困难，已有的方法并不能很好的处理，证明我们的方法在处理更加困难的任务上有很好的效果。

(2) In T1, our CRECL is inferior to RP-CRE because the classification and contrastive network in CRECL have not been fully trained at the beginning. When more tasks are cumulated, CRECL is trained sufficiently, resulting in its superiority over the baselines and less performance drop. Since the catastrophic forgetting becomes more severe on such scenarios, the results imply that CRECL can tackle the catastrophic forgetting better. 
% 2.我们的方法在前面一两轮和同样使用了BERT的方法有轻微差距，这是因为分类网络和对比网络还没有得到充分训练；而在之后几轮任务之中，我们的方法表现性能更优，灾难性遗忘下降更为缓慢，这证明了我们模型的优越性。

(3) All compared models' performance is well on T1, but declines when more new tasks arrive due to more severe catastrophic forgetting.
Compared with EA-EMR, EMAR and CML, the rest models' performance decline is more slight. For example, from the comparison between EMAR (using Bi-LSTM) and EMAR+BERT, we can see that EMAR+BERT's performance decline significantly slows down, proving that BERT helps the model alleviate the catastrophic forgetting better. It is because BERT has good feature discrimination ability and better captures the relevant features, making catastrophic forgetting less severe. 
% 3.所有方法的第一轮准确率都较高，证明所有方法都可以有效地对一个任务进行分类。但随着任务的到来，灾难性遗忘在EA-EMR,EMAR,CML显著发生。而剩下的方法则表现出一定的抵抗灾难性遗忘的特性。例如从EMAR （使用了Bi-LSTM）和EMAR+BERT进行对比，可以看到效果下降有了明显的减缓，这证明预训练语言模型BERT存在着一定的抵抗灾难性遗忘的能力，由于BERT有良好的特征区分能力，其可以更好的捕捉到相关特征，让灾难性遗忘的发生更加不明显。

(4) CREL+ATM's results show that incorporating attention memory module fails to improve CRECL's performance well, because the contrastive network is able to maintain the uniformity and the alignment of data distribution. Thus there is no need for additional attention memory module to help the model refine relation prototypes.
% 4.引入注意力记忆模块并不能很好的改善我们模型的效果，由于对比学习网络本来就存在的保持分布的规整性，所以不需要额外的注意力记忆模块去帮助模型精炼原型。

(5) Compared with the models using metric learning (EA-EMR, EMAR, CML, EMAR+BERT), we adopt InfoNCE loss instead of the margin ranking loss as our loss function. With this loss, CRECL is taught by more negative relation information to understand how to regularize data representation space, resulting in the more alleviation of catastrophic forgetting.

\begin{table*}[t]
%\small
%\setlength\tabcolsep{5.5pt}  
\centering
\caption{Our framework's accuracy (\%) with different memory sizes on FewRel.}\label{table:mem_size}
\vspace{-0.2cm}
\resizebox{0.7\textwidth}{!}{
\begin{tabular}{l|llllllllll}
\hline
CRECL  & T1   & T2   & T3   & T4   & T5   & T6   & T7   & T8   & T9   & T10  \\ 
\hline
$L$=5  & 97.1 & 92.1 & 89.7 & 90.0 & 88.2 & 86.6 & 84.4 & 82.8 & 82.5 & 80.2 \\ 
\hline
$L$=10 & 97.8 & 94.9 & 92.7 & 90.9 & 89.4 & 87.5 & 85.7 & 84.6 & 83.6 & 82.7 \\ 
\hline
$L$=15 & 98.9 & 96.3 & 93.8 & 92.6 & 91.4 & 90.0 & 88.0 & 86.6 & 85.8 & 83.5 \\ 
\hline
$L$=20 & 98.1 & 96.0 & 94.1 & 92.8 & 91.9 & 89.7 & 88.5 & 86.3 & 86.1 & 85.2 \\ \hline
\end{tabular}}
\end{table*}

%\vspace{-0.1cm} 
\subsubsection{Ablation Studies}\label{sec:ablation}
\vspace{-0.1cm}
In order to verify the effectiveness and rationality of our framework's important components (steps), we further conducted a series of ablation experiments. CRECL's ablated variants include:
% 为了理解我们的模型为什么有效，我们尝试了一系列的消融实验分析。
% 下面是我们消融后各个名字的含义。

\noindent\textbf{CRECL-MAG}: It is the variant without the margin loss $\mathcal{L}_{mag}$ in the contrastive network.

\noindent\textbf{CRECL-CL1}: It is the variant without the first training process in the contrastive network.

\noindent\textbf{CRECL-CL2}: It is the variant without the second training process in the contrastive network.

\noindent\textbf{CRECL-CL}: It is the variant only having the classification network. %As the prototypes provided by classification network can not be replaced, we show the role of the classification network by this experiment.
%因为分类模型需要为后续训练提供原型，没有办法直接删去，我们用这个实验来补充分类模型的作用

\noindent\textbf{CRECL-K}: In this variant, the typical instances of each relation are selected at random instead of by K-means algorithm.
% 在选择具有代表性的样例的时候，我们不使用Kmeans,而采取随机选k个数据的策略。

\noindent\textbf{CRECL(C)}: This variant uses the classification network to identify the relation of a test instance instead of the similarity comparison in Eq. \ref{eq:pred}.

Figure \ref{fig:ablation} (a) and (b) display all compared models' accuracy of historical cumulative tasks and current task. Due to space limitation, only the results on TACRED are shown, based on which we have the following analyses. 
% The ablation study results on FewRel are listed in \ref{sec:FewRel}.

(1) CRECL-CL performs very well on current task (subfigure (b)) but performs very poorly on historical tasks (subfigure (a)), showing that it overfits current task and its catastrophic forgetting is very severe. It is because that the classification parameters are always tuned to fit with current task rather than old tasks. It shows that the contrastive network is significant to alleviate catastrophic forgetting.

(2) As more new tasks arrive, CRECL-CL1's performance decline on current task (subfigure (b)) is more obvious than its performance decline on historical tasks (subfigure (a)), because CRECL-CL1 pays more attention to distinguish the different relations in old tasks rather than that in current task. It is due to that the data distributions of all historical tasks are adjusted in the second training process that CRECL-CL1 only has in its contrastive network. Comparatively, CRECL-CL2's performance on historical tasks and current task both declines. It shows that only distinguishing the data distribution of current task from that of old tasks in the first training process of contrastive network, is not adequate to alleviate its catastrophic forgetting. Even worse, such adjusting also harms the accuracy of classifying current task. 
%which proves that the model needs to review the historical data and distinguish it through contrastive learning to alleviate catastrophic forgetting. 

(4) CRECL-K is inferior to CRECL, showing that the randomly selected instances cannot well represent relations as those selected by K-means algorithm. As a result, the data distributions of all tasks cannot be adjusted precisely% to be more distinguishable
, which can not alleviate catastrophic forgetting effectively. In addition, CRECL-K's accuracy on current task is not stable also due to the randomness led by its selection strategy of typical instances.

%(5) We show the result of the classification network after each contrastive training, from \textbf{CRECL(C)}, our model can effectively fit the current task, the contrastive learning training target is different from the classification network training goal, the characteristics of the encoder layer are shared and adjustable to a certain extent, and do not significantly weaken our classification model.
(5) Although the contrastive learning loss $\mathcal{L}_2$ is different from the classification network's loss $\mathcal{L}_1$, and the parameters of the encoding layer are shared, the contrastive network's training processes  hardly weaken the classification network's fitness to current task. Thus CRECL(C) still performs well on current task as shown in Figure \ref{fig:ablation} (b). CRECL-MAG's has a relatively small decline on both current and historical tasks, proving that the margin loss $\mathcal{L}_{mag}$ improves the performance by increasing the gap between the optimal and suboptimal results.
%另一方面，我们给出了分类网络在每次对比训练结束后的效果，从实验来看，我们的模型可以有效地拟合当前任务，可以看到对比学习训练目标虽然和分类网络训练目标不同，但encoder layer的特征是一定程度可共享和可调整的，并没有大幅度减弱我们的分类模型。

\subsubsection{Performance Influence of Memory Size} 
%\vspace{-0.1cm} 

For memory-based CRE methods, the model performance is usually related to the storage capacity of their memory modules. Specifically, we found that previous models' performance is very sensitive to the number of stored typical instances $L$. Recall that in CRECL, a relation prototype is the average of $L$ typical instances' representations. The representational ability of such prototype is less sensitive to $L$ when $L$ exceeds a certain value, resulting in CRECL's performance also less sensitive to $L$, as we emphasized in Section \ref{sec:proto}.  

%approximate distribution representation of the data, and then averages the data K, which causes our model to become less sensitive after the K exceeds a certain amount. So we did experiments with different K as shown 
Figure \ref{fig:mem_diff} displays the performance of CRECL and RP-CRE on TACRED w.r.t. different memory sizes ($L$), where the two compared models' performance in T9 and T10 is specially shown in the subfigure (c). 
% We also list CRECL's performance on FewRel w.r.t. different memory sizes in Table \ref{table:mem_size}. 
It shows that although CRECL's performance also declines when $L$ becomes small, it is more stable and higher than RP-CRE's performance, especially when $L\geq 8$. Such results justifies our claim about CRECL's less sensitivity to $L$. In addition, RP-CRE's performance fluctuation is more obvious, possibly because it re-constructs the attention memory network upon each task, so the different task features are not shared in the network.

\subsubsection{Contrastive Learning's Effectiveness on Refining Data Distributions}
\begin{figure}[!htbp]
%\vspace{-0.1cm} 
\centering
\includegraphics[width=0.6\linewidth]{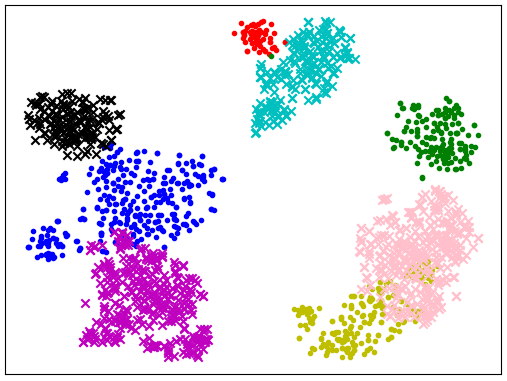}
\vspace{-0.2cm} 
\caption{Data distribution map (better viewed in color) after the contrastive learning in CRECL. Compared with Fig. \ref{fig:res1}, the data distributions of different relations (different colors) are obviously distinguishable, making CRECL classify different relations more easily. %Thus, the model's catastrophic forgetting is alleviated effectively.
}
\label{fig:aftercontrastive}
\vspace{-0.2cm} 
\end{figure}

\vspace{-0.1cm} 
In addition, to investigate the contrastive learning's effects on alleviating catastrophic forgetting, we use \emph{t-SNE} \cite{van2008visualizing} to visualize the data distributions of the same case in Figure \ref{fig:res1} after the training processes of CRECL's classification network and contrastive network. The data distribution map is shown in Figure \ref{fig:aftercontrastive}, which has the same settings of color and point style as Figure \ref{fig:res1}. %As we mentioned in Section \ref{sec:intro}, the data distributions in Figure \ref{fig:res1} are the results after learning the new task with the classification network in CRECL. 
Through comparing these two maps, we find that the data distributions of different relations in the new and old task in Figure \ref{fig:aftercontrastive} are more distinguishable than that in Figure \ref{fig:res1}. Such results are mainly attributed to the prototypical contrastive learning in CRECL on adjusting all data distributions of all tasks, which obviously alleviate CRECL's catastrophic forgetting. It has been also proved by CRECL's superior performance over the baselines displayed in aforementioned experiments. We also note that some yellow dots still intersect the pink crosses, possibly due to the insufficient sampling in the contrastive learning, resulting in less coverage on all relations. We can handle this situation by increasing the batch size.
 
\vspace{-0.2cm} 
\section{Conclusion}
\vspace{-0.2cm}
In this paper, we propose a novel CRE framework, namely CRECL, that consists of a classification network and a contrastive network designed for alleviating the catastrophic forgetting in CRE. Through the prototypical contrastive learning in CRECL, the data distributions of different relations in all tasks are adjusted to be more distinguishable, resulting in CRE performance gains. 
%Our novel model not only effectively alleviates the problem of catastrophic forgetting in CRE task, 
Moreover, CRECL has the ability of class-incremental learning due to its contrasting-based mechanism of achieving relation classification, which is more practical in real-world CRE scenarios than the previous models of classification-based mechanism. Our extensive experimental results demonstrate that CRECL outperforms the SOTA CRE baselines and obtains the best performance on two benchmark datasets. 
% Future work may pay attention to better parameters and some pretrained language model to enhance performance.
\newpage
\bibliography{main-file}

\begin{thebibliography}{29}
\expandafter\ifx\csname natexlab\endcsname\relax\def\natexlab#1{#1}\fi

\bibitem[{Chaudhry et~al.(2018)Chaudhry, Ranzato, Rohrbach, and
  Elhoseiny}]{chaudhry2018efficient}
Arslan Chaudhry, Marc'Aurelio Ranzato, Marcus Rohrbach, and Mohamed Elhoseiny.
  2018.
\newblock Efficient lifelong learning with a-gem.
\newblock \emph{arXiv preprint arXiv:1812.00420}.

\bibitem[{Chen et~al.(2015)Chen, Goodfellow, and Shlens}]{chen2015net2net}
Tianqi Chen, Ian Goodfellow, and Jonathon Shlens. 2015.
\newblock Net2net: Accelerating learning via knowledge transfer.
\newblock \emph{arXiv preprint arXiv:1511.05641}.

\bibitem[{Chen et~al.(2020)Chen, Kornblith, Norouzi, and
  Hinton}]{chen2020simple}
Ting Chen, Simon Kornblith, Mohammad Norouzi, and Geoffrey Hinton. 2020.
\newblock A simple framework for contrastive learning of visual
  representations.
\newblock In \emph{International conference on machine learning}, pages
  1597--1607. PMLR.

\bibitem[{Cui et~al.(2021)Cui, Yang, Yu, Hu, and et~al.}]{cui2021refining}
Li~Cui, Deqing Yang, Jiaxin Yu, Chengwei Hu, and et~al. 2021.
\newblock Refining sample embeddings with relation prototypes to enhance
  continual relation extraction.
\newblock In \emph{Proceedings of the 59th Annual Meeting of the Association
  for Computational Linguistics}, pages 232--243.

\bibitem[{Delange et~al.(2021)Delange, Aljundi, Masana, Parisot, Jia,
  Leonardis, Slabaugh, and Tuytelaars}]{clsurvey1}
Matthias Delange, Rahaf Aljundi, Marc Masana, Sarah Parisot, Xu~Jia, Ales
  Leonardis, Greg Slabaugh, and Tinne Tuytelaars. 2021.
\newblock A continual learning survey: Defying forgetting in classification
  tasks.
\newblock \emph{IEEE Transactions on Pattern Analysis and Machine
  Intelligence}, pages 1--1.

\bibitem[{Devlin et~al.(2019)Devlin, Chang, Lee, and
  Toutanova}]{devlin2018bert}
Jacob Devlin, Ming{-}Wei Chang, Kenton Lee, and Kristina Toutanova. 2019.
\newblock {BERT:} pre-training of deep bidirectional transformers for language
  understanding.
\newblock In \emph{Proceedings of NAACL}, pages 4171--4186.

\bibitem[{Fernando et~al.(2017)Fernando, Banarse, Blundell, Zwols, Ha, Rusu,
  Pritzel, and Wierstra}]{fernando2017pathnet}
Chrisantha Fernando, Dylan Banarse, Charles Blundell, Yori Zwols, David Ha,
  Andrei~A Rusu, Alexander Pritzel, and Daan Wierstra. 2017.
\newblock Pathnet: Evolution channels gradient descent in super neural
  networks.
\newblock \emph{arXiv preprint arXiv:1701.08734}.

\bibitem[{Han et~al.(2020)Han, Dai, Gao, Lin, Liu, Li, Sun, and
  Zhou}]{han2020continual}
Xu~Han, Yi~Dai, Tianyu Gao, Yankai Lin, Zhiyuan Liu, Peng Li, Maosong Sun, and
  Jie Zhou. 2020.
\newblock Continual relation learning via episodic memory activation and
  reconsolidation.
\newblock In \emph{Proceedings of ACL}, pages 6429--6440.

\bibitem[{Han et~al.(2018)Han, Zhu, Yu, Wang, Yao, Liu, and
  Sun}]{han2018fewrel}
Xu~Han, Hao Zhu, Pengfei Yu, Ziyun Wang, Yuan Yao, Zhiyuan Liu, and Maosong
  Sun. 2018.
\newblock Fewrel: A large-scale supervised few-shot relation classification
  dataset with state-of-the-art evaluation.
\newblock \emph{arXiv preprint arXiv:1810.10147}.

\bibitem[{Hassabis et~al.(2017)Hassabis, Kumaran, Summerfield, and
  Botvinick}]{hassabis2017neuroscience}
Demis Hassabis, Dharshan Kumaran, Christopher Summerfield, and Matthew
  Botvinick. 2017.
\newblock Neuroscience-inspired artificial intelligence.
\newblock \emph{Neuron}, 95(2):245--258.

\bibitem[{Khosla et~al.(2020)Khosla, Teterwak, Wang, Sarna, Tian, Isola,
  Maschinot, Liu, and Krishnan}]{khosla2020supervised}
Prannay Khosla, Piotr Teterwak, Chen Wang, Aaron Sarna, Yonglong Tian, Phillip
  Isola, Aaron Maschinot, Ce~Liu, and Dilip Krishnan. 2020.
\newblock Supervised contrastive learning.
\newblock \emph{arXiv preprint arXiv:2004.11362}.

\bibitem[{Kirkpatrick et~al.(2017)Kirkpatrick, Pascanu, Rabinowitz, Veness,
  Desjardins, Rusu, Milan, Quan, Ramalho, Grabska-Barwinska
  et~al.}]{kirkpatrick2017overcoming}
James Kirkpatrick, Razvan Pascanu, Neil Rabinowitz, Joel Veness, Guillaume
  Desjardins, Andrei~A Rusu, Kieran Milan, John Quan, Tiago Ramalho, Agnieszka
  Grabska-Barwinska, et~al. 2017.
\newblock Overcoming catastrophic forgetting in neural networks.
\newblock \emph{Proceedings of the national academy of sciences},
  114(13):3521--3526.

\bibitem[{Kirkpatrick et~al.(2016)Kirkpatrick, Pascanu, Rabinowitz, Veness, and
  et~al.}]{ewc}
James Kirkpatrick, Razvan Pascanu, Neil~C. Rabinowitz, Joel Veness, and et~al.
  2016.
\newblock \href {http://arxiv.org/abs/1612.00796} {Overcoming catastrophic
  forgetting in neural networks}.
\newblock \emph{CoRR}, abs/1612.00796.

\bibitem[{Koch et~al.(2015)Koch, Zemel, Salakhutdinov et~al.}]{koch2015siamese}
Gregory Koch, Richard Zemel, Ruslan Salakhutdinov, et~al. 2015.
\newblock Siamese neural networks for one-shot image recognition.
\newblock In \emph{ICML deep learning workshop}, volume~2, page~0. Lille.

\bibitem[{Li and Hoiem(2016)}]{li2017lwf}
Zhizhong Li and Derek Hoiem. 2016.
\newblock Learning without forgetting.
\newblock \emph{CoRR}, abs/1606.09282.

\bibitem[{Mallya and Lazebnik(2017)}]{packnet}
Arun Mallya and Svetlana Lazebnik. 2017.
\newblock Packnet: Adding multiple tasks to a single network by iterative
  pruning.
\newblock \emph{CoRR}, abs/1711.05769.

\bibitem[{Nayyeri et~al.(2019)Nayyeri, Zhou, Vahdati, Yazdi, and
  Lehmann}]{nayyeri2019adaptive}
Mojtaba Nayyeri, Xiaotian Zhou, Sahar Vahdati, Hamed~Shariat Yazdi, and Jens
  Lehmann. 2019.
\newblock Adaptive margin ranking loss for knowledge graph embeddings via a
  correntropy objective function.
\newblock \emph{arXiv preprint arXiv:1907.05336}.

\bibitem[{Oord et~al.(2018)Oord, Li, and Vinyals}]{oord2018representation}
Aaron van~den Oord, Yazhe Li, and Oriol Vinyals. 2018.
\newblock Representation learning with contrastive predictive coding.
\newblock \emph{arXiv preprint arXiv:1807.03748}.

\bibitem[{Parisi et~al.(2019)Parisi, Kemker, Part, Kanan, and
  Wermter}]{clsurvey2}
German~I. Parisi, Ronald Kemker, Jose~L. Part, Christopher Kanan, and Stefan
  Wermter. 2019.
\newblock \href {https://doi.org/https://doi.org/10.1016/j.neunet.2019.01.012}
  {Continual lifelong learning with neural networks: A review}.
\newblock \emph{Neural Networks}, 113:54--71.

\bibitem[{Rebuffi et~al.(2017)Rebuffi, Kolesnikov, Sperl, and Lampert}]{icarl}
Sylvestre{-}Alvise Rebuffi, Alexander Kolesnikov, Georg Sperl, and Christoph~H.
  Lampert. 2017.
\newblock icarl: Incremental classifier and representation learning.
\newblock In \emph{2017 {IEEE} Conference on Computer Vision and Pattern
  Recognition}, pages 5533--5542.

\bibitem[{Thrun and Mitchell(1995)}]{thrun1995lifelong}
Sebastian Thrun and Tom~M Mitchell. 1995.
\newblock Lifelong robot learning.
\newblock \emph{Robotics and autonomous systems}, 15(1-2):25--46.

\bibitem[{Van~der Maaten and Hinton(2008)}]{van2008visualizing}
Laurens Van~der Maaten and Geoffrey Hinton. 2008.
\newblock Visualizing data using t-sne.
\newblock \emph{Journal of machine learning research}, 9(11).

\bibitem[{Wang and Liu(2021)}]{wang2021understanding}
Feng Wang and Huaping Liu. 2021.
\newblock Understanding the behaviour of contrastive loss.
\newblock In \emph{Proceedings of the IEEE/CVF Conference on Computer Vision
  and Pattern Recognition}, pages 2495--2504.

\bibitem[{Wang et~al.(2019)Wang, Xiong, Yu, Guo, Chang, and
  Wang}]{wang2019sentence}
Hong Wang, Wenhan Xiong, Mo~Yu, Xiaoxiao Guo, Shiyu Chang, and William~Yang
  Wang. 2019.
\newblock Sentence embedding alignment for lifelong relation extraction.
\newblock In \emph{Proceedings of the 2019 Conference of the North {A}merican
  Chapter of the Association for Computational Linguistics: Human Language
  Technologies}, pages 796--806.

\bibitem[{Wolf et~al.(2020)Wolf, Debut, Sanh, and
  et~al.}]{wolf-etal-2020-transformers}
Thomas Wolf, Lysandre Debut, Victor Sanh, and et~al. 2020.
\newblock Transformers: State-of-the-art natural language processing.
\newblock In \emph{Proceedings of EMNLP}, pages 38--45.

\bibitem[{Wu et~al.(2021)Wu, Li, Li, Haffari, Qi, Zhu, and
  Xu}]{wu2021curriculum}
Tongtong Wu, Xuekai Li, Yuan{-}Fang Li, Gholamreza Haffari, Guilin Qi, Yujin
  Zhu, and Guoqiang Xu. 2021.
\newblock Curriculum-meta learning for order-robust continual relation
  extraction.
\newblock pages 10363--10369.

\bibitem[{Yan et~al.(2021)Yan, Xie, and He}]{yan2021dynamically}
Shipeng Yan, Jiangwei Xie, and Xuming He. 2021.
\newblock Der: Dynamically expandable representation for class incremental
  learning.
\newblock In \emph{Proceedings of the IEEE/CVF Conference on Computer Vision
  and Pattern Recognition}, pages 3014--3023.

\bibitem[{Zhang et~al.(2017)Zhang, Zhong, Chen, Angeli, and
  Manning}]{zhang2017tacred}
Yuhao Zhang, Victor Zhong, Danqi Chen, Gabor Angeli, and Christopher~D.
  Manning. 2017.
\newblock Position-aware attention and supervised data improve slot filling.
\newblock In \emph{Proceedings of EMNLP}, pages 35--45.

\bibitem[{Zhao et~al.(2020)Zhao, Xiao, Gan, Zhang, and
  Xia}]{zhao2020maintaining}
Bowen Zhao, Xi~Xiao, Guojun Gan, Bin Zhang, and Shu-Tao Xia. 2020.
\newblock Maintaining discrimination and fairness in class incremental
  learning.
\newblock In \emph{Proceedings of the IEEE/CVF Conference on Computer Vision
  and Pattern Recognition}, pages 13208--13217.

\end{thebibliography}

\bibliographystyle{acl_natbib}

\end{document}